\documentclass[twocolumn,english,amssymb,pra,superscriptaddress]{revtex4-1}

\usepackage{amsmath}
\usepackage{color}
\usepackage{environ}
\usepackage{epstopdf}
\usepackage{balance}

\usepackage{pgfplots}
\usepgfplotslibrary{external}
\pgfplotsset{compat=newest}
\pgfplotsset{plot coordinates/math parser=false}

\usepackage{layouts}

\newcommand{\ket}[1]{| #1 \rangle}

\newcommand{\var}[1] {\mathrm{Var}(#1)}
\newcommand{\cov}[2] {\mathrm{Cov}( #1,  #2 )}

\begin{document}

\title{Estimation of Output Channel Noise for Continuous Variable Quantum Key Distribution}

\author{Oliver Thearle}
\email[E-mail: ]{oliver.thearle@anu.edu.au}
\author{Syed M. Assad}
\author{Thomas Symul}

\affiliation{Centre for Quantum Computation and Communication Technology,
Department of Quantum Science, The Australian National University,
Canberra, ACT 0200, Australia}

\begin{abstract}

Estimation of channel parameters is important for extending the range and increasing the key rate of continuous variable quantum key distribution protocols. We propose a new estimator for the channel noise parameter based on the method of moments. The method of moments finds an estimator from the moments of the output distribution of the protocol. This estimator has the advantage of being able to use all of the states shared between Alice and Bob. Other estimators are limited to a smaller publicly revealed subset of the states. The proposed estimator has a lower variance for high loss channel than what has previously been proposed. We show that the method of moments estimator increases the key rate by up to an order of magnitude at the maximum transmission of the protocol.
\end{abstract}

\maketitle

\section{Introduction}
Quantum Key Distribution (QKD) was proposed in 1984 \cite{BB84} as a solution to the key distribution problem. In this problem Alice wants to share a secret key with a remote party, Bob, but she only has public channels available to her. Alice can solve this problem by encoding the key into quantum states. Alice and Bob can then use quantum mechanics to prove their secret is unconditionally secure from an eavesdropping adversary, Eve \cite{VScarani2009}. The key can then be used later from cryptographic purposes. Continuous Variable (CV) QKD uses the quadrature modulations and measurements of the phase and amplitude quadratures from a bright laser to distribute the shared secret \cite{TRalph1999,FGrosshans2002,ALeverrier2009}. The coherent state protocol with homodyne detection \cite{FGrosshans2002} is commonly used for the study of CV QKD \cite{JLodewyck2007, PJouguet2013}. In this protocol Alice sends Bob a series of randomly displaced vacuum states through an unsecured channel. Bob then measures the received states by either switching between quadratures or measuring both simultaneously. Alice and Bob then estimate a bound on the maximum information that may have been intercepted by Eve in the channel. The optimal attack Eve can make on this protocol has been shown to be a Gaussian collective attack \cite{RRenner2009}. This attack assumes that Eve has access to all information lost in the channel. A bound on Eve's information can be found as a function channel transmission, $T$, and excess channel noise relative to the input, $\xi$. In a practical CV QKD protocol these parameters must be estimated from the shared secret between Alice and Bob. This ensures that the correct bound is found for the final secret key. This is currently done by Alice and Bob publicly revealing a random subset of their initial shared secret \cite{ALeverrier2010,JLodewyck2007}. The number of states revealed can be optimized to give an optimal key rate as discussed in Ref.~\cite{ALeverrier2010,RLaszlo2014,ALeverrier2015}.

In this paper we build on some of the ideas presented in Ref.~\cite{ALeverrier2010}. The authors propose a way of estimating the two parameters  by modeling the protocol using as classical loss channel with additive Gaussian noise:
\begin{equation}
y_i = t x_i + z_i \qquad i=1,2,\dots N.
\label{eq:gausmodel}
\end{equation}  
Here $x_i$ is the data sent by Alice, $y_i$ is Bob's measurement data, $z_i$ is a Gaussian noise term with variance $\sigma^2=1+T\xi$ and mean 0 and $t=\sqrt{T}$. This model is well understood and a maximum likelihood estimator (MLE) exists for both the parameters $t$ and $\sigma^2$. The authors then use these estimators to find the worse case for excess noise transmission to find the final key rate.

We propose to use the method of moments in conjunction with the MLE for $t$ to find an alternative estimator for $\sigma^2$. This estimator has a lower variance than the MLE as $T\to0$. Generally this method produces estimators that are typically worse in terms of minimizing variance and bias than other estimation methods. However it has the advantage of only requiring the public exchange of moments rather than sacrificing part of the key for parameter estimation. The estimator we use in this paper is a function of Alice and Bob's variance and transmission estimates. The variances can be estimated and revealed by Alice and Bob individually using the whole shared secret. Estimation of the transmission still requires some of the shared secret to be revealed. Sharing the variance allows the estimator to use more of the accessible information to decrease the variance of the method of moments estimator, $\hat{\sigma}_\mathrm{MM}^2$, without sacrificing more of the shared secret. The variance of the estimator can be further improved by creating a linear combination of method of moments estimator and the MLE. The resulting estimator is the optimum of the two. 

This paper is organized as follows. Sec. \ref{sec:priorwork} covers prior work on noise estimation and explains the coherent state CV QKD protocol in more detail. In Sec. \ref{sec:MME} we explain the method of moments and show how we arrive at the two estimators.  The variance of the method of moments estimators are then found and compared in Sec.~\ref{sec:perf} to other CV QKD noise estimators and shown to be asymptotically unbiased and in Sec.~\ref{sec:conc} we conclude the paper with a discussion of the method of moments estimator and its effects on the final key rate.

\section{The protocol and model}
\label{sec:priorwork}
In this paper we will consider the coherent state protocol with homodyne detection \cite{FGrosshans2002}. In this protocol Alice prepares $N$ displaced vacuum states, $\ket{q_i +ip_i}$, through phase and amplitude quadrature modulation. The displacements $q_i$ and $p_i$ are both random variables sampled from the normal distribution $\mathcal{N}(0,V_\mathrm{A})$. These states are transmitted to Bob through an unsecure channel with transmission $T$ and excess noise $\xi$. The channel is assumed to be under the control of Eve. Bob will then measure the received states using a homodyne detector switching randomly between the phase and amplitude quadratures. In reference to \eqref{eq:gausmodel} we will take $x_i$ as Alice's modulations and $y_i$ as Bob's measurement outcomes. When Alice uses modulation to prepare the states for Bob it is known as a prepare and measure protocol. These protocols have been shown to have an entanglement based equivalent \cite{FGrosshans2003} which is used for the security analysis \cite{RGarcia2006}. 

Eves optimal attack with finite size effects has been shown to be a Gaussian attack \cite{ALeverrierJune2010}. The collective state between Alice and Bob can be assumed to be Gaussian. For the entanglement based protocol it can be described by the covariance matrix
\begin{equation}
\Gamma=
	\begin{pmatrix}
		\left(V_\mathrm{A} + 1\right) \mathbb{I}_2 & \sqrt{T\left(V_\mathrm{A}^2 + 2V_\mathrm{A}\right)} \sigma_z\\
		\sqrt{T\left(V_\mathrm{A}^2 + 2V_\mathrm{A}\right)}\sigma_z & \left(TV_\mathrm{A}+1+T\xi\right)\mathbb{I}_2
	\end{pmatrix}
\label{eq:cvmatrix},
\end{equation}
where $\sigma_z$ is the Pauli matrix 
\begin{equation}
\sigma_z = \begin{pmatrix}
1 & 0\\
0 & -1
\end{pmatrix}.
\end{equation}
In a prepare and measure scheme Alice and Bob want to find the covariance matrix for the equivalent entanglement based protocol. To do this they reveal a subset of $m < N$ states for estimating the parameter $t$ and $\sigma$. Using the channel model \eqref{eq:gausmodel} for the protocol we have the maximum likelihood estimators \cite{ALeverrier2010}: 
\begin{equation}
\hat{t} = \frac{\sum_{i=1}^m{x_iy_i}}{\sum_{i=1}^m{x_i^2}} \quad \mathrm{and} \quad \hat{\sigma}_{\mathrm{MLE}}^2 = \frac{1}{m} \sum_{i=1}^m{\left(y_i - \hat{t}x_i\right)^2}.
\label{eq:maxest}
\end{equation}
The distributions of these estimators are 
\begin{equation}
\hat{t}\sim\mathcal{N}\left(t,\frac{\sigma^2}{\sum_{i=1}^m{x_i^2}}\right)  \quad \mathrm{and} \quad \frac{m\hat{\sigma}_{\mathrm{MLE}}^2}{\sigma^2}\sim\mathcal{\chi}^2(m-1).
\label{eq:dist}
\end{equation}
As described in Ref.~\cite{ALeverrier2010} the estimates are then used to find the worst case for $t$ and $\sigma^2$. That is the minimum of $t$ and the maximum of $\sigma^2$ with in the confidence interval $1-\epsilon_\mathrm{PE}$. The parameter $\epsilon_\mathrm{PE}$ is the probability that the parameter estimation failed (Typically $\epsilon_\mathrm{PE}=10^{-10}$). Using the theoretical distributions in \eqref{eq:dist} the worst case estimators can then be written as,
\begin{align}
t_{\mathrm{min}} &\approx \hat{t}-z_{\epsilon_\mathrm{PE}/2}\mathrm{Std}(\hat{t})\\ 
\sigma_{max}^2 &\approx \hat{\sigma}_\mathrm{MLE}^2 + z_{\epsilon_\mathrm{PE}/2}\mathrm{Std}(\hat{\sigma}^2_\mathrm{MLE})
\end{align}
Here $\mathrm{Std}$ is the standard deviation function and $z_{\epsilon_\mathrm{PE}/2}=\mathrm{erf}^{-1}(1-\epsilon_\mathrm{PE}/2)$ where $\mathrm{erf(x)}$ is the error function.  We can rewrite \eqref{eq:cvmatrix} for the worst case noise and transmission,
\begin{equation}
\Gamma_{\epsilon_{\mathrm{PE}}} = 
	\begin{pmatrix}
		\left(V_\mathrm{A} + 1\right) \mathbb{I}_2 & t_{\mathrm{min}}\sqrt{V_\mathrm{A}^2 + 2V_\mathrm{A}} \sigma_z\\
		t_{\mathrm{min}}\sqrt{V_\mathrm{A}^2 + 2V_\mathrm{A}}\sigma_z & \left(t^2_{\mathrm{min}}V_\mathrm{A}+\sigma^2_{\mathrm{max}}\right)\mathbb{I}_2.
	\end{pmatrix}
\label{eq:cvgncmatrix}
\end{equation}
Another proposed estimator from Ref.~\cite{RLaszlo2014} uses a second modulation transmitted with the key to assist the estimation of the channel parameters. This assumes the second modulation will experience the same channel as the modulation used for the final key. For the protocol analyzed in their paper, Alice sends Bob squeezed displaced vacuum states with a squeezed quadrature variance of $V_\mathrm{S}$. By setting $V_\mathrm{S}=1$ the protocol becomes the coherent state protocol. The parameters they estimate are the channel transmission $T$ and the excess noise relative to the output $V_\xi=T\xi$. Thanks to the second modulation this estimator is able to use $N$ states for the key and parameter estimation,
\begin{gather}
\hat{T}= \frac{\left(\sum_{i=1}^N x_{\mathrm{M2},i}y_i\right)^2}{\left(NV_\mathrm{M2}\right)^2},\\
\hat{V}_\xi=\frac{1}{N}\sum^N_{i=1}{\left(y_i - \sqrt{\hat{T}}x_{\mathrm{M2},i}\right)^2} - \hat{T}V_\mathrm{A}-1,
\label{eq:estRL}
\end{gather} 
where $x_{\mathrm{M2},i}$ is the displacement of the second modulation from Alice. These estimators were shown to be asymptotically unbiased and to have the variances: 
\begin{gather}
\mathrm{Var}(\hat{T})=\frac{4}{N}T^2\left(2+\frac{V_\mathrm{N}}{TV_\mathrm{M2}}\right), \\ \mathrm{Var}(\hat{V}_\xi)=\frac{2}{N}V_N^2 + V_\mathrm{A}^2\mathrm{Var}(\hat{T}),
\end{gather}
where $V_\mathrm{N}=1+V_\xi+TV_\mathrm{A}$ and $V_\mathrm{M2}$ is the variance of the second modulation. The authors then suggests using the linear combination in Eq.~\eqref{eq:bayesian} to find the optimal estimator $T^\mathrm{opt}$ and $V_\xi^\mathrm{opt}$ at a high channel transmission. 
\begin{equation}
\hat{\theta}_\mathrm{opt}=\alpha\hat{\theta}_1 + (1-\alpha)\hat{\theta}_2,
\label{eq:bayesian}
\end{equation}
where $\hat{\theta}_1$ and $\hat{\theta}_2$ are two different estimators for either $V_\xi$ or $T$. The optimum value of $\alpha$ to achieve a minimum variance from two independent estimators is given by,
\begin{equation}
\alpha = \frac{\mathrm{Var}(\hat{\theta}_2)}{\mathrm{Var}(\hat{\theta}_1) + \mathrm{Var}(\hat{\theta}_2)}.
\label{eq:alpha}
\end{equation}
This can be found by minimising $\mathrm{Var}(\hat{\theta}_{opt})$ with respect to $\alpha$. The variance of $\hat{\theta}_{opt}$ is then given by,
\begin{equation}
\mathrm{Var}(\hat{\theta}_{opt})=\frac{\mathrm{Var}(\hat{\theta_1})\mathrm{Var}(\hat{\theta_2})}{\mathrm{Var}(\hat{\theta_1}) + \mathrm{Var}(\hat{\theta_2})}.
\end{equation}
By construction $\hat{\theta}_{opt}$ will have a variance less than or equal to the variance of the estimators $\hat{\theta}_1$ and $\hat{\theta}_2$. The linear combination will also preserve the bias properties of the two estimators.\\
Once the channel parameters are estimated Alice and Bob will select an appropriate reconciliation protocol and correct the remaining $n=N-m$ states for errors. In this paper we will only consider reverse reconciliation \cite{FGrosshans2002}. Alice and Bob then hash their raw secret key to produce an information-theoretically secure final key \cite{JLodewyck2007}.

The asymptotic key rate for the coherent state protocol with reverse reconciliation is bounded by \cite{JLodewyck2007}
\begin{equation}
K   \geq I(x:y) - S(y:E),
\end{equation}
where $I(x:y)$ is the mutual information between Alice and Bob and $S(E:y)$ is the mutual information Eve has with Bob. Both of these terms can be calculated from the channel parameters. This bound can be rewritten to include the effects for reconciliation efficiency, $\beta$, parameter estimation, $\frac{n}{N}$ and $\epsilon_{\mathrm{PE}}$ on a finite key \cite{ALeverrier2010},
\begin{equation}
K = \frac{n}{N}[\beta I(x:y)-S_{\epsilon_{PE}}(y:E)],
\end{equation}
where $S_{\epsilon_{PE}}(y:E)$ is calculated from the worst case estimates of our channel parameters. The reconciliation efficiency, $\beta$, is related to the amount of information Alice and Bob must sacrifice in order to perform this step. For a given transmission and noise of a channel the choice of reconciliation protocol can be optimized to maximize $\beta$ \cite{PJouguet2011}.\\
\begin{figure}[ht]
\centering
\includegraphics{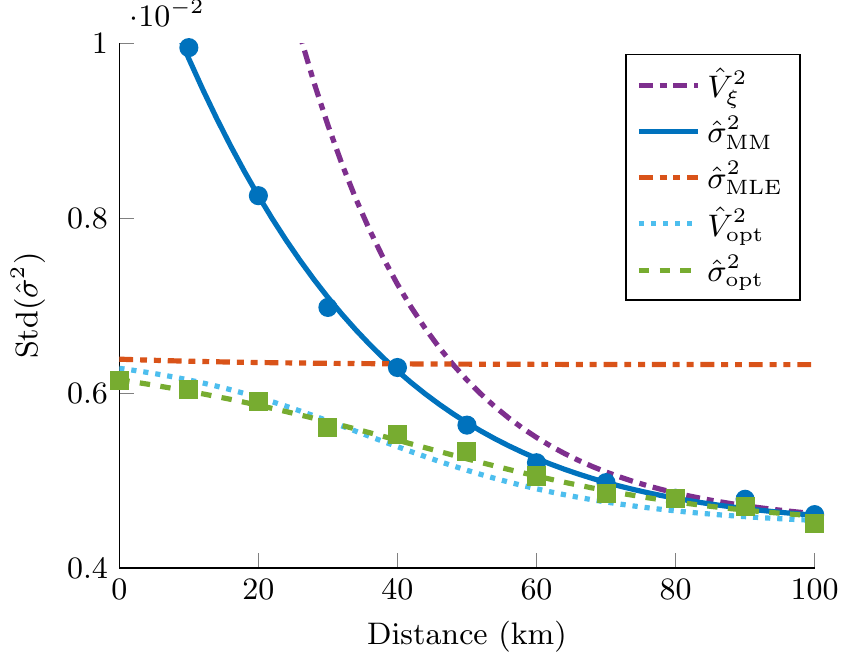}
\caption{\label{fig:std} (Color online) Plot of the standard deviation of the different noise estimators vs distance in a fiber channel: $\hat{V}_\xi^2$ (dot dashed), $\hat{\sigma}_\mathrm{MM}^2$ (solid), $\hat{\sigma}^2_\mathrm{MLE}$ (dot dot dashed), $\hat{V}_{\xi}^\mathrm{opt}$ (dotted) and $\hat{\sigma}_{\mathrm{opt}}^2$ (dashed) with simulations of $\hat{\sigma}_\mathrm{MM}^2$ (circles) and $\hat{\sigma}_\mathrm{opt}^2$ (squares). The parameters used were $V_A = 3$, $\xi = 0.01$, $m = 0.5{\times}10^5$, $N=10^5$, $\beta=0.95$ and $V_2 = 10$.}
\end{figure}
\section{The Method of Moments Estimator}
\label{sec:MME}
The estimators in Eq.~\eqref{eq:maxest} are found by maximizing the log likelihood probability function $\ln p(x_i,y_i;\sigma^2,t,V_A)$. An alternative is to use the method of moments \cite{SKay1993} to find the estimators. The method of moments is a simple way to find an estimator but it has no optimality properties. It performs best with a long data record which makes it suitable to CV QKD as typically the data record is $>10^8$ \cite{PJouguet2013}. To use the method we first find a probability distribution describing our observations in terms of the parameters we want to estimate. In the case of Bob's measurements the distribution is given by $\mathcal{N}\left(0,t^2V_\mathrm{A}+\sigma^2\right)$. The moments of this distribution can then be solved as a system of equations for the parameters we want to estimate. As Bob's data is normally distributed around 0 the first moment will be zero and the second moment is given by the variance,
\begin{equation}
\sigma_\mathrm{B}^2=t^2 V_\mathrm{A} + \sigma^2.
\label{eq:sigmaB}
\end{equation}
All other moments for this distribution will be a function of $\sigma_\mathrm{B}^2$ giving only one independent non zero moment. This allows us to only find one estimator. We are most interested in maximizing the keyrate for long distance CV QKD. The limiting factor for protocols with a high loss channel is the excess noise \cite{ALeverrier2009}. For this reason we willl concentrate on finding a better estimator for the output noise. The variance in Eq. \eqref{eq:sigmaB} can be used to estimate $t$ but the process is made more difficult by requiring an estimate of $\sigma^2$. Starting with Eq. \eqref{eq:sigmaB} and substituting the estimator for $t$ and the sample variance for $\sigma_\mathrm{B}$ we find an initial estimator for the noise relative to the output.
\begin{equation}
	\hat{\sigma}_\mathrm{mm}^2=\hat{\sigma}_\mathrm{B}^2-\hat{t}^2 V_\mathrm{A},
	\label{eq:sigma2}
\end{equation}
where $\hat{\sigma}_\mathrm{B}^2$ is given by the sample variance $\frac{1}{N}\sum{y_i^2}$. To use this estimator Alice and Bob can publicly reveal $V_\mathrm{A}$ and $\sigma^2_\mathrm{B}$ without giving away any more of the shared secret to Eve \cite{ALeverrier2010}. A decrease to the variance of $\hat{\sigma}^2_\mathrm{mm}$ can be made by using the estimated second moment of Alice's collective state, $\hat{\sigma}^2_\mathrm{A}$. The variances of $\hat{\sigma}^2_\mathrm{mm}$ and the new estimator $\hat{\sigma}^2_\mathrm{MM}$ are compared in appendix \ref{apd:varmm}. This improvement comes from increasing the covariance between $\sigma_\mathrm{B}^2$ and $t^2\sigma_\mathrm{A}^2$ and is demonstrated by the following property of variance,
\begin{multline}
\mathrm{Var}(\sigma_\mathrm{B}^2-t^2\sigma_\mathrm{A}^2)=\mathrm{Var}(\sigma_\mathrm{B}^2) + \mathrm{Var}(t^2V_\mathrm{A}) \\- 2\mathrm{Cov}(\sigma_\mathrm{B}^2,t^2V_\mathrm{A}) .
\label{eq:thvar}
\end{multline}
Substituting $\hat{\sigma}_\mathrm{A}=\frac{1}{N}\sum{x_i^2}$ we arrive at our final method of moments estimator,
\begin{align}
\hat{\sigma}_\mathrm{MM}^2 &= \hat{\sigma}_B^2-\hat{t}^2\hat{\sigma}_A^2.
\end{align}
Using Eq. \eqref{eq:thvar} we can already see the improvement in the variance of the estimator $\hat{\sigma}_\mathrm{MM}^2$ will have over $\hat{\sigma}_\mathrm{MLE}^2$ as the transmission approches zero for a fixed value of $V_\mathrm{A}$ and $m$. We find for both estimators their variance is given by the variance of the variance of Bob's measurements, 
\begin{equation}\var{\hat{\sigma}_\mathrm{B}^2} = \frac{2\sigma_\mathrm{B}^4}{M},\end{equation}
where $M$ is the number of samples used to find $\hat{\sigma}_\mathrm{B}^2$. Taking the varaince of both estimators we find $\var{\hat{\sigma}_\mathrm{MM}^2}$ is better by a factor of $\frac{m}{N}$.\\
An interesting point is $\hat{\sigma}_\mathrm{MM}^2 = \hat{\sigma}_\mathrm{MLE}^2$ when both estimators are used on the $N$ transmitted states. Such as the case at the range limit of a protocol where we reveal almost all of the states for parameter estimation for a positive key. Using \eqref{eq:maxest} on the $N$ transmitted states we find
\begin{align}
\hat{\sigma}_\mathrm{MLE}^2 &= \frac{1}{N} \sum_{i=1}^N{(y_i - \hat{t}x_i)^2} \label{eq:sums}\\
			&= \frac{1}{N} \sum_{i=1}^N{y_i^2} - \frac{1}{N} \frac{\left(\sum_{i=1}^N{x_iy_i}\right)^2}{\sum_{i=1}^N{x_i^2}} \\
			&= \hat{\sigma}_\mathrm{B}^2-\hat{t}^2 \hat{\sigma}_\mathrm{A}^2\\
			&=\hat{\sigma}_\mathrm{MM}^2.
\label{eq:eqiv}
\end{align}
If in Eq. \eqref{eq:sums} we split the summation into the two subsets, the publically revealed $m$ states and the secret $n$ states, we can also show
\begin{align}
\hat{\sigma}_\mathrm{MM}^2 &= \frac{1}{N} \sum_{i=1}^m{(y_i - \hat{t}x_i)^2} + \frac{1}{N} \sum_{i=1}^n{(y_i - \hat{t}x_i)^2}\\
&=\frac{1}{N}(m\hat{\sigma}_\mathrm{MLE}^2 + n\hat{\sigma}_\mathrm{MM''}^2),
\end{align}
where $\hat{\sigma}_\mathrm{MM''}^2$ is the method of moments estimator for the noise applied to the $n$ states to be used to generate the final key. It is easy to show that $\hat{\sigma}^2_{\mathrm{MLE}}$ and $\hat{\sigma}^2_{\mathrm{MM''}}$ are independent estimators given $\hat{t}$ and $\hat{\sigma}^2_{\mathrm{MLE}}$ are also independent \cite{ALeverrier2009}. This leads to the next estimator we present in this paper. As in Ref.~\cite{RLaszlo2014} we can find an optimum linear combination of our two estimators. Using \eqref{eq:bayesian} we find an optimum estimate of the noise,
\begin{equation}
\hat{\sigma}_\mathrm{opt}^2 = \alpha \hat{\sigma}_\mathrm{MLE}^2 + (1-\alpha)\hat{\sigma}_\mathrm{MM''}^2.
\end{equation}
Here $\alpha$ is given in Eq.~\eqref{eq:alpha}.
\begin{figure}[ht]
\centering
\includegraphics{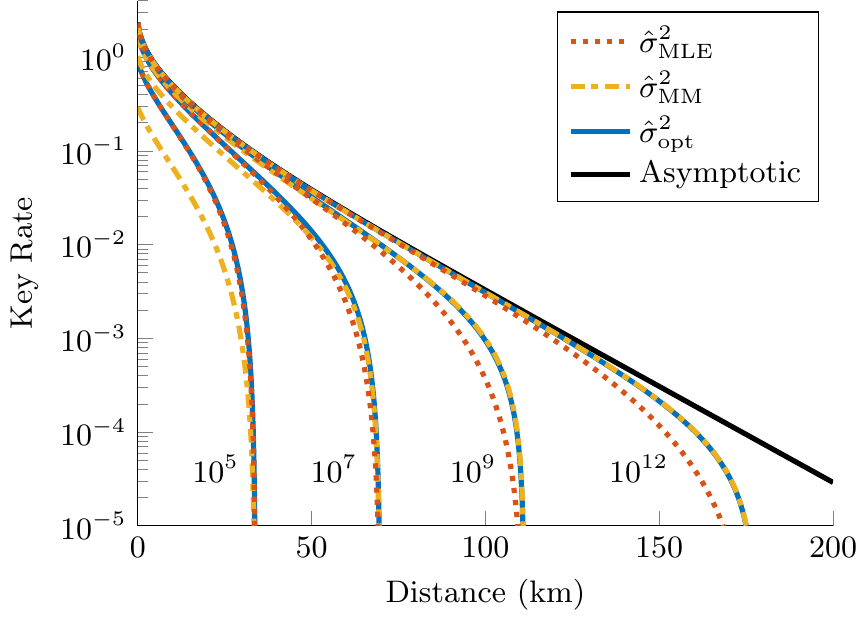}
\caption{\label{fig:keyrate} (Color online) Plot of key rate with finite key effects relating to parameter estimation. The values $V_\mathrm{A}$ and $m$ have been optimized  with $\xi=0.01$ to maximize key rate using $\hat{\sigma}_\mathrm{MLE}^2$ (dotted), $\hat{\sigma}_\mathrm{MM}^2$ (dot dashed) and $\hat{\sigma}_\mathrm{opt}^2$ (solid) to estimate the excess noise for (from left to right) $N=10^5$, $N=10^7$, $N=10^9$ and $N=10^{12}$. The asymptotic key rate with $V_\mathrm{A}$ optimized is also plotted (black solid).}
\end{figure}
\section{Performance}
\label{sec:perf}
For the purposes of CV QKD it is important to consider the variance and the bias of the parameter estimators. Finding an unbiased estimator with a minimized variance will ultimately lead to an increase in the key rate and secure distance of the protocol.\\
For the method of moments estimators the variance and mean are difficult to find due to the division required for $\hat{t}$. For this paper we use a standard method in uncertainty analysis where the variance is approximated from a first order Taylor series expansion \cite{SKay1993}. Given an estimator $\hat{\theta}$ that is some function of $\boldsymbol J = \{ J_1(\boldsymbol y),J_2(\boldsymbol y),\dots,J_r(\boldsymbol y)\}$, where $J_i(\boldsymbol y)$ is some statistic from the data vector $\boldsymbol y$, we find the variance is approximated by,
\begin{equation}
\var{\hat{\theta}(\mathbf{J})} \approx \frac{\partial \hat{\theta}}{\partial \mathbf{J}}\bigg|_{\mathbf{J}=\boldsymbol\mu}^T \boldsymbol C_{J}\: \frac{\partial \hat{\theta}}{\partial \mathbf{J}}\bigg|_{\mathbf{J}=\boldsymbol\mu}
\label{eq:vartheta}
\end{equation}
and the mean by
\begin{equation}
E(\hat{\theta}(\mathbf{J}))\approx\hat{\theta}(\boldsymbol\mu).
\label{eq:meantheta}
\end{equation}
Here $\boldsymbol\mu$ is the expected value of our statistics $\boldsymbol J$ and $\boldsymbol C_{J}$ is the covariance matrix for $\boldsymbol J$. This method assumes that the statistics $\boldsymbol J$ will have a low variance and the estimator $\hat{\theta}$ will be roughly linear around $\boldsymbol \mu$. That is Eq.~\eqref{eq:vartheta} and Eq.~\eqref{eq:meantheta} will be the asymptotic variance and mean. To apply this method we rewrite our estimators in terms of the data statistics,  $\hat{\sigma}^2_\mathrm{B}$, $\hat{\sigma}^2_\mathrm{A}$, $\hat{\sigma}^{\phantom{{}+{}}}_{\mathrm{A'B'}}$ and $\hat{\sigma}^2_\mathrm{A'}$. The estimator $\hat{\sigma}_\mathrm{MM}^2$ becomes
\begin{equation}
\hat{\sigma}_\mathrm{MM}^2=\hat{\sigma}^2_\mathrm{B} - \frac{\hat{\sigma}^{\phantom{{}+{}}}_{\mathrm{A'B'}}}{\hat{\sigma}^2_\mathrm{A'}}
\hat{\sigma}^2_\mathrm{A}
\end{equation}
Here $\hat{\sigma}^{\phantom{{}+{}}}_{\mathrm{A'B'}}=\frac{1}{m}\sum{x_iy_i}$ is the covariance of the collective state shared by Alice and Bob. We use $\mathrm{A'}$ and $\mathrm{B'}$ to indicate the statistic was estimated from the $m$ subset of states used for parameter estimation. The matrix $\boldsymbol C_J$ can be found using the variance of the sample variance and the properties of the covariance and variance functions. The elements of $\boldsymbol C_J$ are given in appendix \ref{apd:cjMM}. Applying Eq.~\eqref{eq:vartheta} the variance is given by
\begin{equation}
\label{eq:varMM}
\mathrm{Var}(\hat{\sigma}_\mathrm{MM}^2)\approx\frac{2\sigma^4}{N} + \left(\frac{1}{m} - \frac{1}{N} \right)4t^2\sigma^2V_A.
\end{equation}
The final variance in \eqref{eq:varMM} was achieved by making the substitution $\sigma_\mathrm{A}=\sigma_\mathrm{A'}=V_\mathrm{A}$, $\frac{\sigma_\mathrm{A'B'}}{\sigma_\mathrm{A'}}=t$ and $\sigma_\mathrm{B}^2=t^2 V_\mathrm{A} + \sigma^2$. For the estimator $\hat{\sigma}_\mathrm{MM''}^2$ we find a similar equation, 
\begin{equation}
\label{eq:var}
\mathrm{Var}(\hat{\sigma}_\mathrm{MM''}^2)\approx\frac{2\sigma^4}{n} + \left(\frac{1}{m} + \frac{1}{\sigma^2n} \right)4t^2\sigma^2V_A.
\end{equation}
As $\hat{\sigma}_\mathrm{MM''}^2$ uses different statistics we will have a new $\boldsymbol C_J$. This is given in  appendix \ref{apd:cjMMdashdash}. The variance of the optimal estimator is given by \cite{RLaszlo2014}
\begin{equation}
\mathrm{Var}(\hat{\sigma}_\mathrm{opt}^2) = \frac{\mathrm{Var}(\sigma_\mathrm{MLE}^2)\mathrm{Var}(\sigma_\mathrm{MM''}^2)}{\mathrm{Var}(\sigma_\mathrm{MLE}^2)+\mathrm{Var}(\sigma_\mathrm{MM''}^2)}.
\label{eq:optvar}
\end{equation}
The standard deviation of the estimators $\hat{\sigma}_\mathrm{MM}^2$ and $\hat{\sigma}_\mathrm{opt}^2$ are plotted as a function of the channel distance in Fig.~\ref{fig:std}. Finding the expected value our estimators using Eq. \ref{eq:meantheta} shows the estimator,  $\hat{\sigma}_\mathrm{MM}^2$ is asymptotically unbiased.\\
With a simulation of the coherent state protocol using $N=10^5$ we show good agreement in Fig.~\ref{fig:std} with Eq.~\eqref{eq:var} and Eq.~\eqref{eq:optvar}. In practical demonstrations $N$ has been of the order $10^8$ to $10^9$ \cite{PJouguet2013}.
\begin{figure}
\centering
\includegraphics{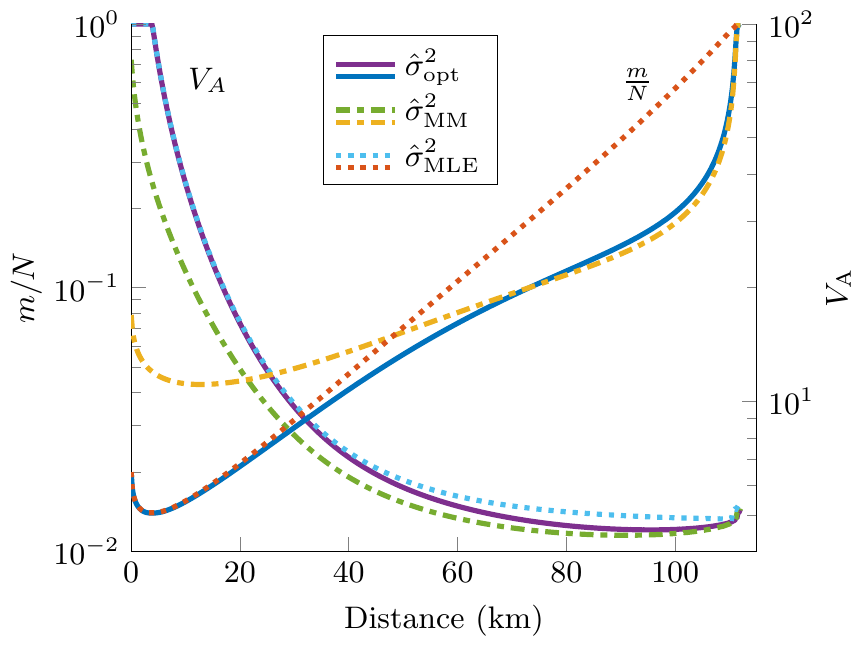}
\caption{\label{fig:Varopt} (Color online) The optimized values of $\frac{m}{N}$ and $V_\mathrm{A}$ for the key rates in Fig.~\ref{fig:keyrate} where $N=10^9$ using $\hat{\sigma}_\mathrm{MLE}^2$ (dotted), $\hat{\sigma}_\mathrm{MM}^2$ (dot dashed) and $\hat{\sigma}_\mathrm{opt}^2$ (solid) to estimate the excess noise.} 
\end{figure}
\section{Discussion and Conclusion}
\label{sec:conc}
We have investigated using the method of moments in place of a maximum likelihood estimator to estimate the channel noise, $\sigma^2$ for CV QKD protocols. We found the method of moments produced a similar estimator to the MLE for $\sigma^2$. The method of moments estimator was found to have a lower variance with high loss channels. By combining the method of moments estimator with the MLE we showed that we could optimise the variance for any value of channel loss. The improvement in the variance allows the protocol to use more of the shared secret for key generation. This is demonstrated in Fig~\ref{fig:Varopt}. These estimators were shown to have a similar performance to the estimators proposed in Ref.~\cite{RLaszlo2014}. The advantage of the estimators proposed in our paper is they do not require additional modulations and only use the states used for key generation and less for parameter estimation. \\
The improvement in variance of $\hat{\sigma}_\mathrm{MM}^2$ and $\hat{\sigma}_\mathrm{opt}^2$ over the MLE comes from using more of the available information to Bob for estimation. This decrease in variance will lead to an increase in the key rate but not necessarily an increase in maximum distance as we show in Fig.~\ref{fig:keyrate}. The reason for this comes back to Eq.~\eqref{eq:eqiv}. We show in Fig.~\ref{fig:Varopt} that as $T\to0$ the optimal $m$ goes to $N$ which leads to both estimators approaching $\hat{\sigma}^2_\mathrm{MLE}$. We note that our optimum estimator produces a key rate that is always greater than or equal to key rates found with the other estimators in Fig.~\ref{fig:keyrate} as expected.\\ 
With simplicity of the method of moments this estimator can also be modified to be used with other CV QKD protocols such as the four state protocol \cite{ALeverrier2009} or to include more protocol parameters \cite{PJouguet2012}.
\begin{acknowledgments}
We wish to thank to Andrew Lance and Matthew James for useful discussions leading to this work. This research is supported by the Australian Research Council (ARC) under the Centre of Excellence for Quantum Computation and Communication  Technology  (CE110001027) and Linkage Project (LP130100783).
\end{acknowledgments}

\bibliography{bibfile}

\appendix

\section{Variance of $\sigma^2_\mathrm{mm}$}
\label{apd:varmm}
Using the same method described in Sec. \ref{sec:perf} we find the variance for $\sigma^2_\mathrm{mm}$ is given by
\begin{equation}
\mathrm{Var}(\hat{\sigma}_\mathrm{mm}^2)\approx\frac{2\sigma^4}{N} + \frac{2t^4V_A^2}{N} + \left(\frac{1}{m} - \frac{1}{N} \right)4t^2\sigma^2V_A.
\label{eq:thing}
\end{equation}
Here the covariance $\boldsymbol C_J$ can be found using appendix \ref{apd:cjMM} and setting the appropriate values to 0. With Eq. \ref{eq:thing} we find
\begin{equation}
\mathrm{Var}(\hat{\sigma}_\mathrm{mm}^2)=\frac{2t^4V_A^2}{N} +  \mathrm{Var}(\hat{\sigma}_\mathrm{MM}^2).
\end{equation}
This agrees with our claim that $\mathrm{Var}(\hat{\sigma}_\mathrm{mm}^2)>\mathrm{Var}(\hat{\sigma}_\mathrm{MM}^2)$.

\section{Elements of $\boldsymbol C_J$}
\label{apd:cj}

\subsection{$\boldsymbol C_J$ for $\hat{\sigma}_\mathrm{MM}^2$}
\label{apd:cjMM}
The diagonal terms for the covariance matrix $\boldsymbol C_J$ for the estimator $\hat{\sigma}_\mathrm{MM}$ are given by,
\begin{align}
\var{\hat{\sigma}_\mathrm{A}^2} &= \frac{2 \sigma_\mathrm{A}^4}{N}, & \var{\hat{\sigma}_\mathrm{A'}^2}& = \frac{2 \sigma_\mathrm{A'}^4}{m},\nonumber\\
\var{\hat{\sigma}_\mathrm{B}^2} &= \frac{2 \sigma_\mathrm{B}^4}{N}  & \mathrm{and}\quad \var{\hat{\sigma}_\mathrm{A'B'}} &= \frac{1}{m}(2 t^2 \sigma_\mathrm{A'}^4 + \sigma ^2 \sigma_\mathrm{A'}^2).\nonumber
\end{align}
The off diagonal terms are given by,
\begin{align}
\cov{\hat{\sigma}_\mathrm{A}^2}{\hat{\sigma}_\mathrm{B}^2} &= 2 t^2 \frac{{\sigma}_\mathrm{A}^4}{N}, & \cov{\hat{\sigma}_\mathrm{A}^2}{\hat{\sigma}_\mathrm{A'}^2} & = 2 \frac{{\sigma}_\mathrm{A'}^4}{N}, \nonumber\\
\cov{\hat{\sigma}_\mathrm{A}^2}{\hat{\sigma}_\mathrm{A'B'}} &= 2 t \frac{\sigma_\mathrm{A'}^4}{N}, & \cov{\hat{\sigma}_\mathrm{A'}^2}{\hat{\sigma}_\mathrm{A'B'}} & = 2 t \frac{\sigma_\mathrm{A'}^4}{m},\nonumber\\
\cov{\hat{\sigma}_\mathrm{B}^2}{\hat{\sigma}_\mathrm{A'}^2} &= 2t^2 \frac{\sigma_\mathrm{A'}^4}{N}&\nonumber
\end{align}
and
\begin{align}
\cov{\hat{\sigma}_\mathrm{B}^2}{\hat{\sigma}_\mathrm{A'B'}} &= 2 t \frac{t^2 \sigma_\mathrm{A'}^4 +  \sigma^2\sigma_\mathrm{A'}^2}{N}.\nonumber
\end{align}

\subsection{$\boldsymbol C_J$ for $\hat{\sigma}_\mathrm{MM''}^2$}
\label{apd:cjMMdashdash}
The diagonal terms for the covariance matrix $\boldsymbol C_J$ for the estimator $\hat{\sigma}_\mathrm{MM''}$ are given by,
\begin{align}
\var{\hat{\sigma}_\mathrm{A''}^2} &= \frac{2 \sigma_\mathrm{A''}^4}{n}, & \var{\hat{\sigma}_\mathrm{A'}^2}& = \frac{2 \sigma_\mathrm{A'}^4}{m},\nonumber\\
\var{\hat{\sigma}_\mathrm{B''}^2} &= \frac{2 \sigma_\mathrm{B''}^4}{n}  & \mathrm{and}\quad \var{\hat{\sigma}_\mathrm{A'B'}} &= \frac{1}{m}(2 t^2 \sigma_\mathrm{A'}^4 + \sigma^2 \sigma_\mathrm{A'}^2)\nonumber
\end{align}
The off diagonal terms are given by,
\begin{align}
\cov{\hat{\sigma}_\mathrm{A''}^2}{\hat{\sigma}_\mathrm{B''}^2} &= 2 t^2 \frac{\hat{\sigma}_\mathrm{A''}^4}{N}, & \cov{\hat{\sigma}_\mathrm{A''}^2}{\hat{\sigma}_\mathrm{A'}^2} & = 0, \nonumber\\
\cov{\hat{\sigma}_\mathrm{A''}^2}{\hat{\sigma}_\mathrm{A'B'}} &= 0, & \cov{\hat{\sigma}_\mathrm{A'}^2}{\hat{\sigma}_\mathrm{A'B'}} & = 2 t \frac{\sigma_\mathrm{A'}^4}{m},\nonumber\\
\cov{\hat{\sigma}_\mathrm{B''}^2}{\hat{\sigma}_\mathrm{A'}^2} &= 0 \quad \mathrm{and}& \cov{\hat{\sigma}_\mathrm{B''}^2}{\hat{\sigma}_\mathrm{A'B'}^2} &= 0.\nonumber
\end{align}
Here we use $\mathrm{A''}$ and $\mathrm{B''}$ to indicate the statistic was calculated using the $n$ subset of states used for generating the final key.

\end{document}